# Joint spectral amplitude analysis of SPDC photon pairs in a multimode ppLN ridge waveguide

Ramesh Kumar,[1] Joyee Ghosh[1,*]

[1]*Department of physics, Indian Institute of Technology Delhi, New Delhi 110016*
*E-Mail: *joyee@physics.iitd.ac.in*

In this paper, we study the possible parametric down conversion processes in a periodically poled customized Lithium Niobate (LiNbO3) ridge waveguide. Our analysis of spontaneous parametric down-conversion (SPDC), first, with a Gaussian pump beam mode and second, with an anti-symmetric Hermite-Gaussian HG (1,0) pump beam mode predict the possible down conversion processes in each case. From our *JSA* analysis, it is evident that the generated photons pairs in all these cases are negatively correlated and have orthogonal polarizations. In case of the former, degenerate photon pairs are emitted at 1550 nm with the highest efficiency in the fundamental waveguide mode. While, in case of the latter, non-degenerate photon pairs in different higher order spatial modes are generated. Such photons, thus, have multiple degrees of freedom, like polarization and spatial modes, which can be further harnessed towards hyper-entangled photons for quantum information applications.

## I. INTRODUCTION

The progress in quantum optics and quantum communication depends crucially on compact sources of entangled photon pairs. Spontaneous parametric down-conversion (SPDC) is used as one of the most popular resources for their generation. Typically, the detection of an idler photon heralds the presence of a signal photon to be used in subsequent experiments [1]. These down-converted photons share certain non-classical correlations and can be entangled in different degrees of freedom, providing numerous applications in the above fields [2]. Thus, the ability to control the characteristics of quantum states of light has become increasingly important [3]. In this regard, the joint spectral amplitude (*JSA*) analysis of the emitted photon pairs in SPDC can play a crucial role in studying and controlling these characteristics.

The Joint spectral amplitude (*JSA*) characterizes the structure and joint spectrum of the generated photons. From *JSA* we can extract information about the different degrees of freedom of the photon pairs and their quantum correlations. The shape and profile of the pump beam used in SPDC can influence the structure of the photon pairs [4]. Typically, a laser cavity supports different types of transverse modes in the form of Hermite-Gaussian (HG) modes. HG beams retain their beam shape and spherical phase fronts as they propagate through free space and have rectangular symmetry along the propagation axis. Apart from the fundamental HG (0,0) the most common higher order mode corresponds to HG (1,0) which is an anti-symmetric mode and whose field extends over a large distance in the transverse plane and has a sign variation in the phase. Such beams can be also generated using a spatial light modulator. We have studied the influence of such a pump beam in the *JSA* of the generated photons through a quantum mechanical analysis of SPDC in a waveguide.

Although the first experiments where performed using bulk crystals, PDC in bulk suffers low generation rates and collection efficiencies. Thus, in the past years, PDC in nonlinear waveguides have accumulated substantial interest. A waveguide channels light through the material, giving rise to an enhanced down-conversion rate and a collinear propagation of all involved fields in well-defined spatial modes. Guided wave SPDC sources perform better due to tight lateral confinement of the waveguide modes [1], which on one hand, allows such sources to be pumped at reduced power levels. While on the other hand, they still achieve high photon fluxes in comparison to what bulk crystals can achieve. In recent years ridge waveguides have attracted a lots of attention because they offer large bandwidth of operation and have better lateral confinement of light as compared to rectangular or strip waveguides [5]. A ridge structure particularly promising high-power handling capacity, which is essential for in nonlinear optical processes in general [6-8].

In general, a waveguide supports several spatial modes for the propagating pump, signal and idler photons, corresponding to their respective wavelengths. Due to confinement of the propagating pump, signal and idler in waveguide not only are their spatial characteristics modified but also their spectral properties are affected. The energy distribution between the signal and idler photons, and hence their wavelengths, is governed by the phase matching of the longitudinal components of the wave vector. Each set of three spatial modes pump, signal and idler leads to a unique spectrum corresponding to a particular PDC process and such spectra superimpose at the output of waveguide [1].

In this paper, we present a theoretical analysis of PDC in a waveguide studying the possible spatial mode structure of the generated photons. In particular, we have considered a 5% MgO- doped ppLN ridge waveguide which is type II phase-matched and customized for twin photons generation at 1550 nm following the SPDC process: 775 nm → 1550 nm +1550 nm. Ridge waveguides in ppLN are less sensitive to photo-refraction and therefore, can be operated at an higher optical powers without degradation due to optical damage [9]. We predict the possible down conversion processes first with a Gaussian pump beam and then with a HG (1,0) pump beam. Further a joint spectral analysis ensures that the waveguide parameters are ambient for generating frequency correlated photons. In Section II, we consider the theoretical aspects of SPDC in a waveguide. In Section III, we predict the possible down conversion processes with a Gaussian pump beam. In Section IV, we predict and study the possible down conversion processes with an antisymmetric or HG (1,0) input pump beam. We summarize and conclude in Section V.

## II. SPDC IN A WAVEGUIDE; THEORETICAL CONSIDERATION AND WAVEGUIDE CHARACTERISTICS

The parametric down conversion (PDC) joint photon state can be written as:

$$|\psi\rangle = B \sum_{lmn} A_p^l A_{lmn} \iint d\omega_s d\omega_i f_{lmn}(\omega_s, \omega_i) \times \hat{a}_s^{(m)\dagger}(\omega_s) \hat{a}_i^{(n)\dagger}(\omega_i)|0,0\rangle. \quad (1)$$

Where $B$ is an overall constant, $l$, $m$ and $n$ are the modes of pump, signal and idler, respectively inside the waveguide. $A_p^l$ is the overlap integral as in Eq. (2) of the input beam field distribution $E_{in}^{pump}(\mathbf{r})$ with the pump mode of the waveguide at $\lambda_p$, with $u_p^{(l)}$ as the normalized field profile of the pump mode and defined as:

$$A_p^{(l)} = \iint d\mathbf{r}\, u_p^{(l)}(\mathbf{r}) E_{in}^{pump}(\mathbf{r}), \quad (2)$$

with $\mathbf{r} \equiv (x, y)$. If we explicitly consider the spatial modes propagating inside the guided material, it is seen that the generated biphoton state is emitted into a superposition of interacting mode triplets ($lmn$). Due to the spatial overlap and interaction between the pump, signal and idler fields, given by the overlap integral $A_{lmn}$, each such mode triplet exhibits a different overall down-conversion efficiency. $A_{lmn}$ is the spatial overlap of the three interacting modes $l$, $m$ and $n$, defined as:

$$A_{lmn} = \int_A d\mathbf{r}\, u_p^{(l)}(\mathbf{r}) u_s^{(m)}(\mathbf{r}) u_i^{(n)}(\mathbf{r}). \quad (3)$$

We consider the joint spectral amplitude of the two photons defined as:

$$f_{lmn}(\omega_s, \omega_i) = \alpha(\omega_s + \omega_i)\, \phi_{lmn}(\omega_s, \omega_i). \quad (4)$$

Where $\alpha(\omega_s + \omega_i) = e^{-\left(\frac{\omega_s + \omega_i - \omega_p}{\sigma_p}\right)^2}$ is the pump envelope function, which we consider to be a Gaussian function incorporating the frequencies involved in the PDC process that satisfy the conservation of energy. $\phi_{lmn}(\omega_s, \omega_i)$ is the phase matching function which is defined as:

$$\phi_{lmn}(\omega_s, \omega_i) = \mathrm{sinc}\left[\Delta\beta_{lmn}(\omega_s, \omega_i)\frac{L}{2}\right] \exp\left(i\Delta\beta_{lmn}(\omega_s, \omega_i)\frac{L}{2}\right). \quad (5)$$

Where $\Delta\beta_{lmn}(\omega_s, \omega_i) = \beta_p^{(l)}(\omega_s + \omega_i) - \beta_s^{(m)}(\omega_s) - \beta_i^{(n)}(\omega_i) - \beta_{QPM}$ is the phase matching condition in the

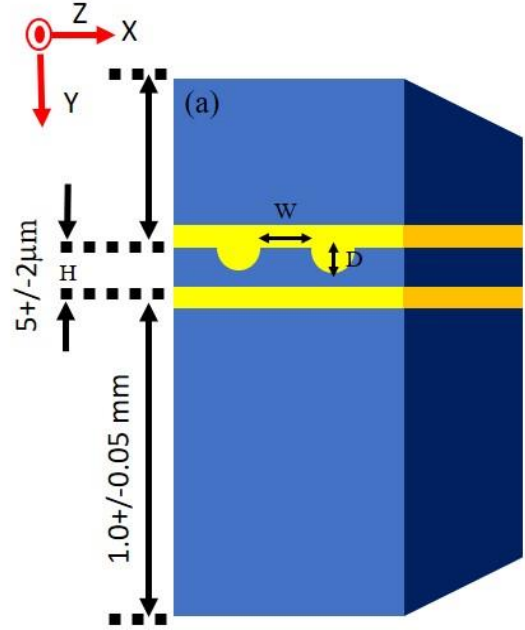

Fig. 1 (a) Schematic of the ridge waveguide, reproduced from HC-Photonics; (b) Fundamental mode profile supported by the ridge waveguide at pump wavelength 775 nm.

waveguide corrected by the $\beta_{QPM}$ quasi phase matching vector. Here $\beta_{p,s,i}^{(\tau)} = \frac{2\pi}{\lambda_{p,s,i}} n_{p,s,i}^{(\tau)}$ is the wave vector of the pump, signal and idler, respectively corresponding to the effective indices $n_{p,s,i}^{(\tau)}$ of the same inside the waveguide, where $\tau = (l, m, n)$ modes.

Among many nonlinear optical materials, Lithium Niobate (LN) is a versatile $\chi^{(2)}$ medium with a large effective nonlinear coefficient and can be obtained as waveguides [2]. We theoretically study the possible down-conversion processes through type II SPDC in a 5% MgO-doped congruent periodically-poled LN ridge waveguide. For the same we have considered a modal analysis of the ppLN ridge waveguide in our previous work [10], whose dimensions are W = 6 μm, H = (5 ± 2) μm, and D = 2.8 μm. In Fig. 1(a) we have shown the schematic of the ridge waveguide for the PDC process. In Fig 1(b) we have shown the fundamental (0,0) mode profile supported by the ridge waveguide at a pump wavelength 775 nm. For calculating the refractive index of LN for different polarizations and wavelengths of light we have used the relevant Sellmeier equation [11, 12].

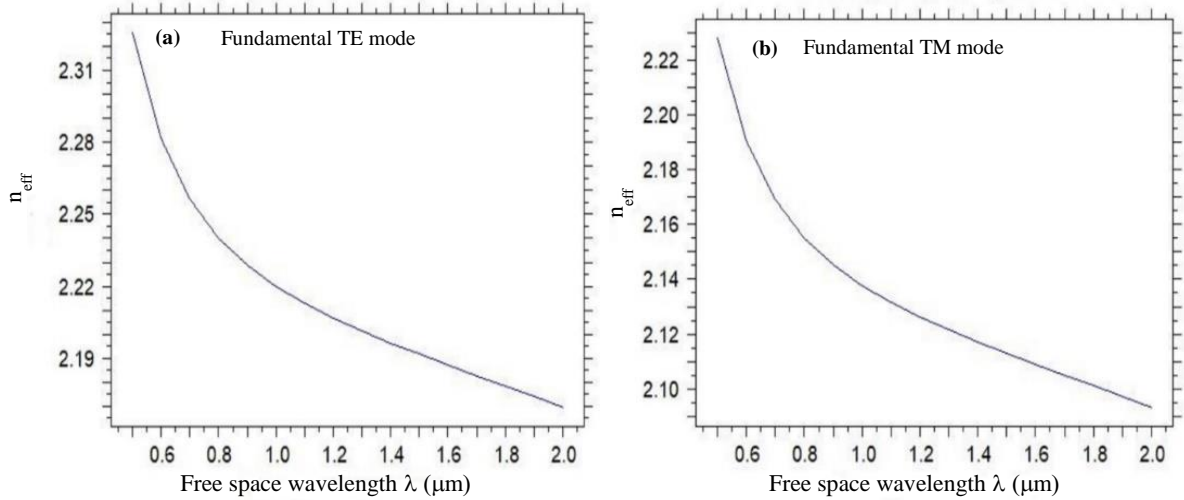

Fig. 2: Variation of the effective index of the waveguide with free space wavelength for the fundamental (a) TE mode and (b) TM mode.

In Fig. 2, we show the variation of effective index of the waveguide with wavelength for the fundamental TE and TM mode. This variation also depends on the shape and size of the waveguide. We have optimized the height of ridge waveguide for generation of photon pair in the telecom wavelength. For all the analysis we have taken the optimized height of the ridge waveguide H= 4.04 μm. The waveguide has large refractive index contrast, $\Delta n \approx 0.8$ and has a large cross section, supporting many higher order modes of the waveguide.

In our type II PDC process x-polarized pump photons near 775 nm are down-converted to x and y-polarized signal and idler photons around 1550 nm. The spatial modes propagating in this waveguide architecture are computed with a numerical model of the dielectric waveguide. Fig.3, shows the first four x-polarized spatial field modes at 1550 nm.

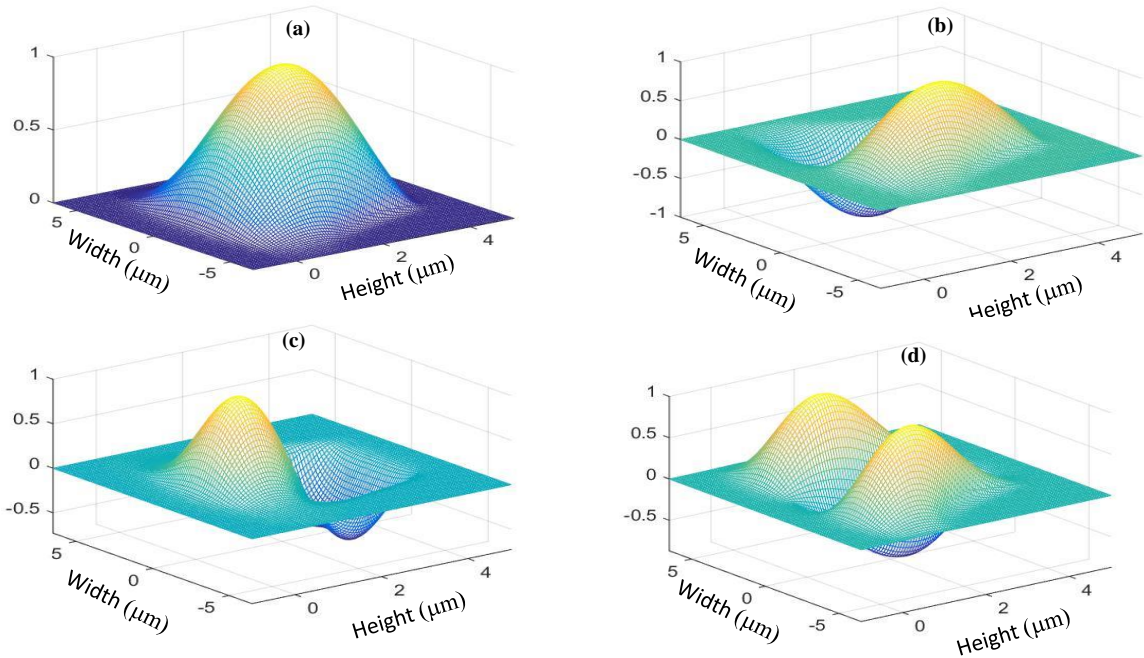

Fig 3: First four TE spatial modes (a) (0,0) (b) (1,0) (c) (0,1) (d) (2,0) at 1550 nm propagating in the ridge waveguide.

## III. POSSIBLE DOWN CONVERSION PROCESSES WITH A GAUSSIAN PUMP BEAM

We first consider an input pump with Gaussian field distribution

$$E_{in}^{pump}(\mathbf{r}) = e^{-\left(\frac{x^2}{a^2} + \frac{y^2}{b^2}\right)} \quad (6)$$

where $2a$ and $2b$ are the x-width and y-width of the Gaussian beam cross section, respectively, at which the field amplitude falls to 1/e of its axial value. For example, one can assume a pulsed laser source acting as the pump with a Gaussian output (bandwidth $\sigma_p$ = 250 GHz) and having an overlap of $A_p^l \approx$ 96% with the fundamental mode of the waveguide [10]. With this we have computed the spatial overlap integral $A_{lmn}$ and the generated wavelengths of a first few dominant processes with a fundamental pump mode are given in Table I.

**TABLE I: Computed values of the generated wavelengths and $A_{lmn}$ for different mode conversions.**

| $E^{(p)}(l_x, l_y) \rightarrow E^{(s)}(m_x, m_y) + E^{(i)}(n_x, n_y)$ | | | | | $\lambda_s$ (nm) | $\lambda_i$ (nm) | $A_{lmn}$ |
|---|---|---|---|---|---|---|---|
| (0,0) | → | (0,0) | + | (0,0) | 1550.0 | 1550.0 | 0.2647 |
| (0,0) | → | (1,0) | + | (1,0) | 1323.9 | 1869.2 | 0.2034 |
| (0,0) | → | (2,0) | + | (2,0) | 1063.8 | 2854.7 | 0.1755 |

We have used the poling period $\Lambda$ = 8.29 μm, pump wavelength $\lambda_p$ = 775 nm, and length of the 5% MgO-doped ppLN waveguide $\mathscr{L}$ = 1 cm, for all simulations in this paper. Below, Table I indicates that the emitted signal/idler photon pairs in this waveguide are degenerate at 1550 nm being emitted in the fundamental mode, and highly non-degenerate at higher order modes of the waveguide. Several other possible down-conversion processes (not shown) are suppressed as they cannot be seen above the noise level in the measurement.

In Fig. 4(a) we have plotted the pump envelope intensity (*PEI*) for the fundamental mode of the waveguide whose width is directly proportional to the bandwidth of the pump beam. *PEI* signifies the conservation of the energy in the SPDC process. In Fig. 4(b), we have plotted the phase matching intensities (*PMI*) for the three dominating processes (in Table I,) with a fundamental pump mode for the output wavelength range of 550 nm. The linewidth of the *PMI* is inversely proportional to the length of the waveguide. *PMI* gives an information of the conservation of momentum. The yellow (bright) lines in the *PMI* shows that $\Delta\beta_{lmn}(\omega_s, \omega_i) = 0$. The slope of the *PMI* curves in Fig. 4(b) of these three processes is negative in each case, which indicate that the generated photons in different modes are negatively correlated [10]. The *PMI* have different value of the slopes corresponding to different processes. This is due to the different group velocities at different frequencies. Instead of being straight lines, the *PMIs* are slightly curved as shown for the relevant wavelength range of 550 nm, due to dispersion in the waveguide, which is low in our case. When the radius of curvature of the *PMI* is low it corresponds to a high dispersion, and vice versa.

In Fig. 5, we showed the joint spectral intensity (*JSI*) that gives information of the probability density of the emitted photons at $\lambda_s$ and $\lambda_i$. The *JSI* plots show that although our waveguide is suitable for generating degenerate photon pairs at 1550 nm as the dominant process but it can also generate photons at different wavelengths elaborated in Table 1 due to other SPDC process. These are detailed in Fig. 5(a) and (b) which shows the predicted signal and idler wavelength, respectively using the *JSI* of the emitted joint photon states. The *JSI* in general signifies that energy and momentum are conserved in the relevant SPDC process leading to the generation of photon pairs in particular wavelengths. There are more possible signal/idler modes generated at higher wavelengths of idler corresponds to lower wavelengths of signal (not shown in fig. 5(a)-(b)). The strength of the *JSI* peak depends on the value of $A_{lmn}$, which is the spatial overlap of the three interacting modes *l*, *m* and *n* corresponding to pump, signal and idler as in Eq. (3).

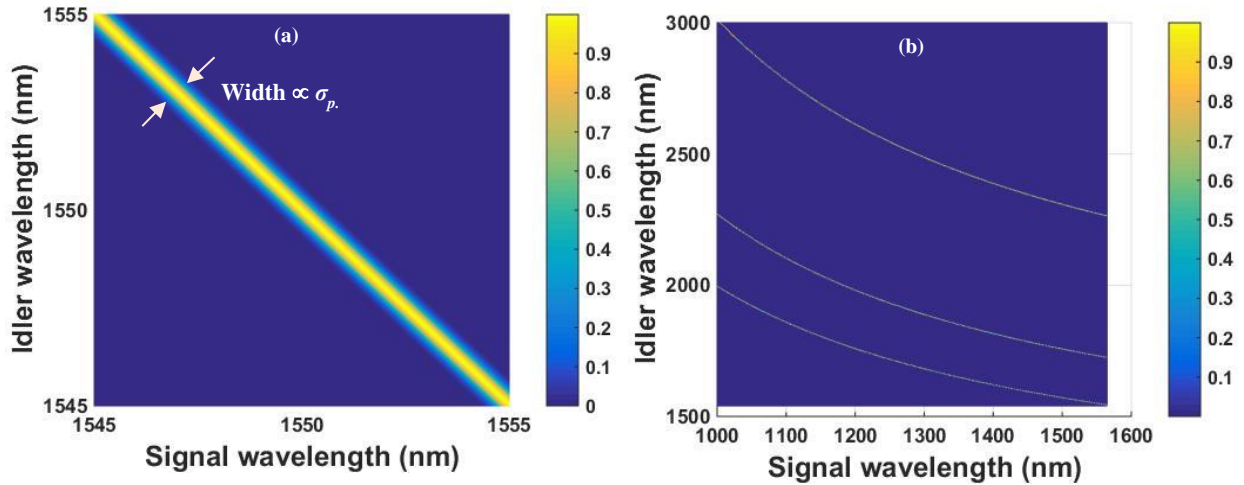

Fig. 4 (a) Pump envelope intensity for a gaussian pump mode; (b) Phase matching intensities of different modes with a gaussian pump mode.

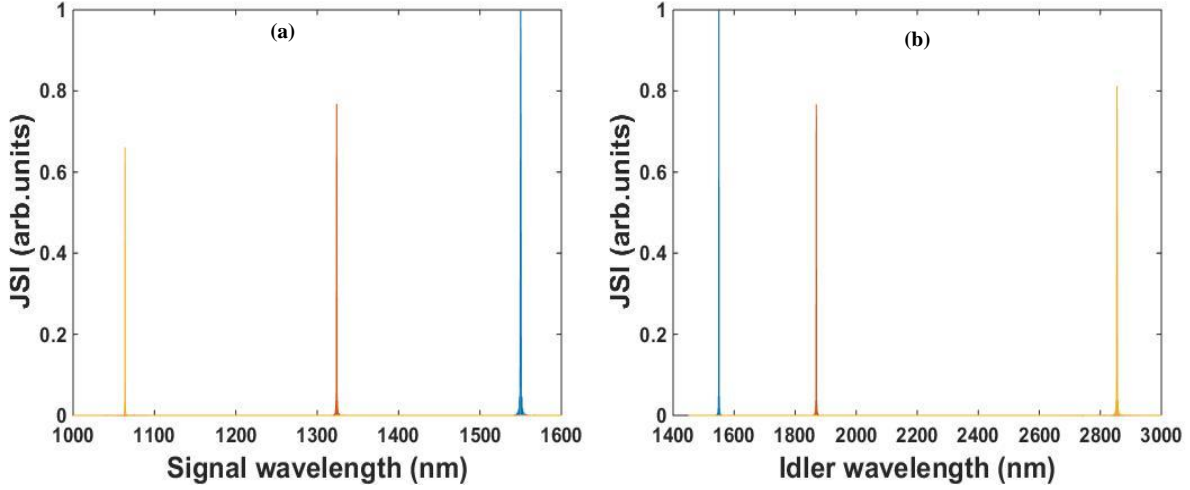

Fig.5: *JSI* versus predicted wavelength of (a) signal, and (b) idler.

## IV. POSSIBLE DOWN CONVERSION PROCESSES WITH A HERMITE-GAUSSIAN HG (1,0) PUMP BEAM

In the previous section we have simulated our waveguide using a Gaussian pump beam according to Eq. (6). In this section we simulate our waveguide with a Hermite-Gaussian HG (1,0) pump beam mode. HG (1,0) whose input field $E_{in}^{pump}(\mathbf{r})$ defined as:

$$E_{in}^{pump}(\mathbf{r}) = H_m\left(\frac{\sqrt{2}x}{a}\right) H_n\left(\frac{\sqrt{2}y}{b}\right) exp\left[-\left(\frac{x^2}{a^2} + \frac{y^2}{b^2}\right)\right]. \quad (7)$$

Where $H_\zeta$ is a Hermite polynomial of order $\zeta = (m,n)$. $(a,b)$ is the optimum $(x,y)$ spot size of the pump beam. The launching position of the HG (1,0) pump beam in the waveguide is aligned in such a way that most of the energy goes into the (1,0) mode of the waveguide. In this way we have optimized the input beam parameters, given in the Table II.

**TABLE II: Optimal input beam parameters**

| Parameter | Optimum value ($\mu m$) |
|---|---|
| Launch position (X-direction) | 0 |
| Launch position (Y-direction) | 1.6 |
| Gaussian X-width ($2a$) | 5 |
| Gaussian Y-width ($2b$) | 2.5 |

For the same input pump beam, we have calculated the overlap integral $A_p^{(l)} = \iint d\mathbf{r}\, u_p^{(l)}(\mathbf{r}) E_{in}^{pump}(\mathbf{r})$ with various field profiles of the normalized pump mode $u_p^{(l)}$ of the waveguide given in Table III. As expected, the overlap integral of the input HG (1,0) beam with the (1,0) mode of the waveguide is maximum as compared to other modes supported by the waveguide. We have computed the signal/idler wavelengths and corresponding $A_{lmn}$ for the different mode conversions with the (1,0) mode of the waveguide given in Table IV. Here one must note that the input HG (1,0) beam given in Eq. (7), has an overlap of 97.5 % with the (1,0) mode of the waveguide. This should be incorporated with the $A_{lmn}$ in Table IV for an overall amplitude in each component of the PDC state in Eq. (1).

**TABLE III: Overlap integrals of various waveguide modes with an input HG (1,0) profile.**

| Mode number | Overlap integral |
|---|---|
| (1,0) | 0.975 |
| (0,0) | $4.68 \times 10^{-8}$ |
| (0,1) | $6.63 \times 10^{-8}$ |
| (2,0) | $7.47 \times 10^{-7}$ |

Table IV shows the four main dominating SPDC processes possible with an HG (1,0) mode as the input pump beam in the given waveguide. Here, it should be noted that signal photons at 1550 nm could also be obtained (while the idler is at some other wavelength) by changing the height/width of the waveguide. For the waveguide dimensions considered as the same in Sec. III, the possible signal/idler wavelengths emitted are detailed in Table. IV.

**TABLE IV: Computed values of the generated wavelengths and $A_{lmn}$ for different mode conversions.**

| $E^{(p)}(l_x,l_y) \rightarrow E^{(s)}(m_x,m_y) + E^{(i)}(n_x,n_y)$ | | | | | $\lambda_s$ (nm) | $\lambda_i$ (nm) | $A_{lmn}$ |
|---|---|---|---|---|---|---|---|
| (1,0) | → | (0,0) | + | (1,0) | 1484.3 | 1621.8 | 0.2034 |
| (1,0) | → | (1,0) | + | (0,0) | 1494.4 | 1609.9 | 0.2046 |
| (1,0) | → | (2,0) | + | (1,0) | 1244.3 | 2054.9 | 0.1443 |
| (1,0) | → | (1,0) | + | (2,0) | 1220.3 | 2123.8 | 0.1382 |

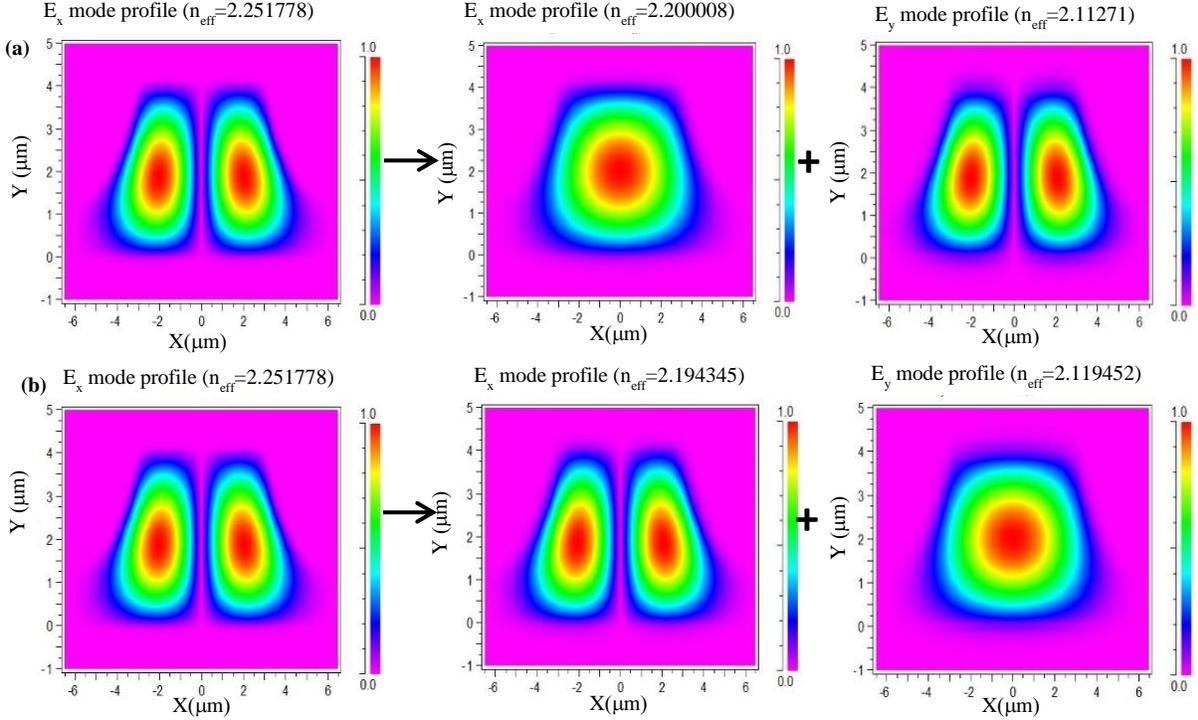

Fig.6: (a) Mode conversions to different orders of the signal-idler spatial modes in type II PDC process : (a) (1,0)→ (0,0) + (1,0) and (b) (1,0) → (1,0) + (0,0), showing the first two processes of Table. IV.

The spatial overlap of the three interacting modes ($lmn$) corresponding to the pump, signal and idler respectively, can be obtained from Eq.(3) as before. We calculated the coupling constants (in terms of these overlap integrals $A_{lmn}$) of these mode conversion processes to estimate the efficiency in each case. Correspondingly, in Fig. 6, we display the different normalized mode profiles pertaining to the different signal and idler wavelengths and polarizations generated in the first two possible SPDC processes, elaborated in Table IV, using a HG (1,0) pump beam. It may be noted that these two processes are almost equally probable, corresponding to similar values of $A_{lmn}$. They also correspond to similar wavelengths for the non-degenerate signal and idler photons.

In Fig. 7(a), we have plotted the phase matching intensity (*PMI*) corresponding to different conversion processes with a HG (1,0) pump beam, which depicts the conservation of momentum in a PDC process. The slopes of the all the *PMI* curves with HG (1,0) pump beam are also negative as that in Sec. 3. with a Gaussian pump beam. This indicates that the generated photon pairs in different mode conversion processes have negative correlation. For the first two PDC processes given in Table IV, we are able to generate the signal and idler photons in a combination of symmetric and antisymmetric modes. If we represent the '0' as symmetric mode and '1' as anti-symmetric mode, then the joint state can be written as:

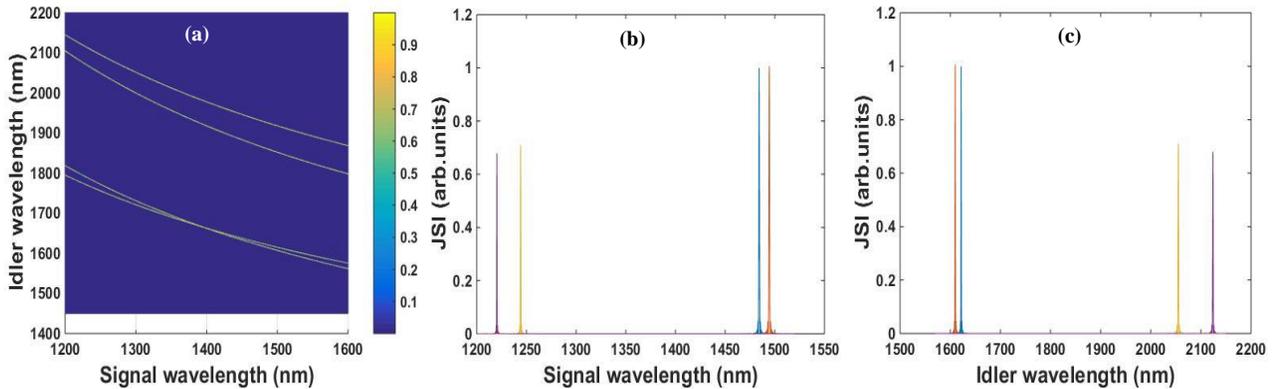

Fig 7: (a) *PMI* of different modes; (b) predicted signal wavelength vs *JSI*; (c) predicted idler wavelength vs *JSI*, all with HG (1,0) pump mode.

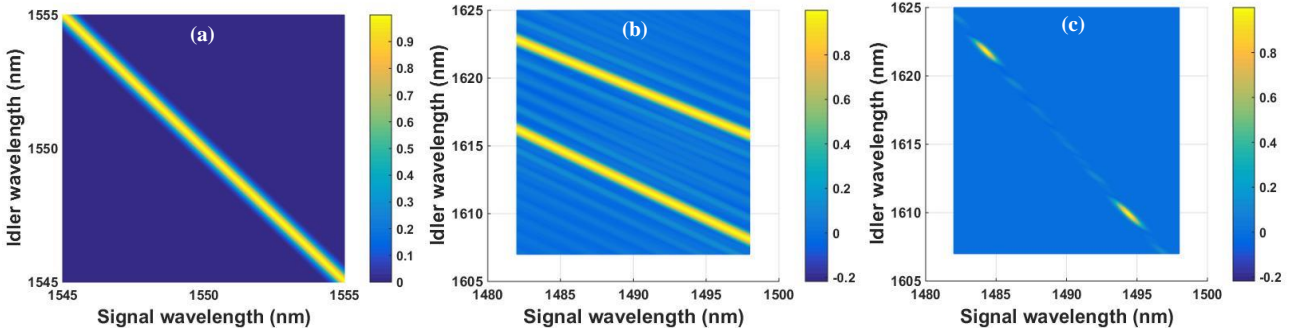

Fig.8: (a) HG (1,0) *PEF*; (b) *PMF* and (c) *JSA* of the first two processes in Table 4 corresponding to a HG (1,0) pump mode.

$$|H_s 0_s, V_i 1_i\rangle + |H_s 1_s, V_i 0_i\rangle = |H_s, V_i\rangle \otimes (|0_s, 1_i\rangle + |1_s, 0_i\rangle). \quad (8)$$

Where $H_s$ and $V_i$ indicate signal and idler with horizontal and vertical polarizations, respectively. In Fig 7 (a), we can predict the mode entanglement in an output collection band width of (1390 nm-1403) for the signal and (1661 nm-1668 nm) for the idler.

In Fig. 7 (b), (c) we have plotted the *JSI* with the HG (1,0) pump mode with a value of the overlap integral $A_p^{(l)}$ taken from the Table III. With the help of *JSI*, we are able to predict the probability of generated photons frequency. In Fig. 7(b) we show the predicted signal wavelengths while in Fig. 7(c) we show the predicted idler wavelengths, which can be measure above the noise level. We have plotted the pump envelope function *PEF* in Fig. 8 (a) and the phase matching function (*PMF*) of first two PDC processes (elaborated in Table IV.) with a HG (1,0) input pump beam mode of the waveguide is plotted in Fig 8(b). The lower bright line in Fig. 8(b) corresponds to the first PDC process of Table IV and the upper bright line corresponds to the second PDC process of Table IV. The direction of *PMF* is tilted at an arbitrary angle $\theta$ with the horizontal axis. This angle $\theta$ can be determined [1] as follow:

$$\tan \theta = -\frac{n_p^g(\omega_p) - n_s^g(\omega_s)}{n_p^g(\omega_p) - n_i^g(\omega_i)}, \quad (9)$$

Where $n_\mu^g(\omega_\mu)$, $\mu = p, s, i$ is the group index of pump, signal and idler. In the first and second processes of Table IV, the *PMF* is tilted at $\theta \approx -28.7°$, and $\theta \approx -28.85°$ respectively which is shown in Fig 8(b). Each triplet possesses a unique spectrum $f_{lmn}(\omega_s, \omega_i)$ which is defined in Eq (4). In Fig 8(c), we show the joint spectral amplitude of these first two PDC processes with (1,0) mode of the waveguide. The resulting two photon spectrum is again negatively correlated due to the negative slope of the phase matching function. In the first process of Table IV, non-degenerate photon pairs are generated at $\lambda_s = 1484.3$ nm and $\lambda_i = 1621.8$ nm, while in the second process of Table IV, non-degenerate photon pairs are generated at $\lambda_s = 1494.4$ nm and $\lambda_i = 1609.9$ nm which is also depicted in their *JSA*, in Fig. 8(c). Although the signal/idler bandwidths for these two processes do not explicitly have an overlapping region in the current context, one can slightly change the waveguide dimensions to achieve this overlapping bandwidth. If such a case is feasible, one can try to fabricate a waveguide with cascaded poling periods or bi-periods $\Lambda_1$ and $\Lambda_2$ for Type II phase-matching, in such a manner that $\Lambda_1$ supports the PDC process: $p^H \to s^H + i^V$ (as considered in our study) and in addition, $\Lambda_2$ supports another PDC process: $p^H \to s^V + i^H$. Then for the latter, similar to Eq. (8) one can write the joint entangled state of the photons as

$$|V_s 0_s, H_i 1_i\rangle + |V_s 1_s, H_i 0_i\rangle = |V_s, H_i\rangle \otimes (|0_s, 1_i\rangle + |1_s, 0_i\rangle). \quad (10)$$

Now following the states in Eqs. (8) and (10), being generated together in the same waveguide and considering all possibilities, one can re-write the joint photon state to be

$$(|H_s 0_s, V_i 1_i\rangle + |H_s 1_s, V_i 0_i\rangle) + (|V_s 0_s, H_i 1_i\rangle + |V_s 1_s, H_i 0_i\rangle) = \\ (|H_s, V_i\rangle + |V_s, H_i\rangle) \otimes (|0_s, 1_i\rangle + |1_s, 0_i\rangle), \quad (11)$$

which is a hyper-entangled state of the photon pairs, being entangled both in polarization and spatial/modal degree of freedom. This has immense applications in the field of quantum communication and quantum information in general.

## V. DISCUSSION

We have predicted and studied the possible down conversion processes in a multimode ppLN ridge waveguide for two cases, with a Gaussian pump beam (elaborated in Table I) and a HG (1,0) pump beam (elaborated in Table IV). Our study indicates that degenerate photon pairs are emitted in the fundamental mode of the waveguide at 1550 nm with the highest efficiency for the first case. In addition, other SPDC processes leading to the emission of non-degenerate photon pairs at different wavelengths (with some at higher order modes) also occur in the same. In particular, SPDC with a HG (1,0) pump beam shows that the efficiency of the emitted photons in each of these different modes can be enhanced by choosing appropriate input pump beam parameters. The generated photons with a HG (1,0) pump beam are also negatively correlated like those with a Gaussian pump beam. The advantage of using a HG (1,0) beam as the pump is the generation of non-degenerate photons pairs at different modes with higher efficiency that can be used in quantum photonics applications.

Due to the conservation of parity, a symmetric pump mode (in case of a Gaussian beam) considered for the PDC processes (detailed in Table I), can only generate down-converted signal-idler pairs both in either symmetric modes or anti-symmetric modes. In these processes we cannot generate signal and idler photons in a combination of symmetric and anti-symmetric modes. On the other hand, with an anti-symmetric pump mode, like HG (1,0), detailed in Table IV) we can generate the signal and idler photons in a combination of symmetric and anti-symmetric modes. Such generated photons can have multiple degrees of freedom such as polarization, mode and frequency and hence can be applied towards the study of hyper-entangled photons. With proper tailoring of the photons further these properties can enhance quantum information tasks in quantum communication and photonic quantum computing. Non-degenerate photons generated with a HG (1,0) pump beam mode can also be applied as a source of heralded single photons. Based on the detection of an idler photon generated in an antisymmetric mode of the waveguide corresponding to a particular wavelength, the partner photon can act like a single photon emitted in a symmetric mode and specifically for same cases in the fundamental (0,0) mode of the waveguide, which is the further scope in our work.

Our investigations revealed an intricate interplay between the spatial, spectral and polarization degree of freedom, which enables the design of spatially single-mode PDC sources, e.g. for the heralding pure single photon states and can be harnessed to generate photon states hyper-entangled in the spatial and polarization degree of freedom.

## ACKNOWLEDGEMENT

We thank the Department of Science & Technology, Government of India for a research grant: DST/ICPS/QuST/Theme-l/2019 (Capital), Project #9.